\documentclass[12pt]{elsarticle}

\usepackage{graphics}
\usepackage{graphicx}
\usepackage{amsmath}
\usepackage{amssymb}
\usepackage{bm}

\newcommand{\rar}{\rightarrow}

\newcommand{\ttt}{\texttt}
\newcommand{\mbb}{\mathbb}
\newcommand{\mbf}{\mathbf}
\newcommand{\mca}{\mathcal}

\newtheorem{proposition}{Proposition}

\begin{document}

\journal{Journal of Informetrics and published}

\begin{frontmatter}

\title{Exposing Multi-Relational Networks to Single-Relational Network Analysis Algorithms\footnote{Rodriguez M.A., Shinavier, J., ÒExposing Multi-Relational Networks to Single-Relational Network Analysis Algorithms,Ó Journal of Informetrics, volume 4, number 1, pages 29-41, ISSN:1751-1577, Elsevier, doi:10.1016/j.joi.2009.06.004, LA-UR-08-03931, December 2009.}}

\author[a1]{Marko A. Rodriguez}
\author[a2]{Joshua Shinavier}

\address[a1]{T-5 Center for Nonlinear Studies, Los Alamos National Laboratory, Los Alamos, New Mexico 87545}
\address[a2]{Tetherless World Constellation, Rensselaer Polytechnic Institute, Troy, New York, 12180}

\begin{abstract}
Many, if not most network analysis algorithms have been designed specifically for single-relational networks; that is, networks in which all edges are of the same type. For example, edges may either represent ``friendship," ``kinship," or ``collaboration," but not all of them together. In contrast, a multi-relational network is a network with a heterogeneous set of edge labels which can represent relationships of various types in a single data structure. While multi-relational networks are more expressive in terms of the variety of relationships they can capture, there is a need for a general framework for transferring the many single-relational network analysis algorithms to the multi-relational domain. It is not sufficient to execute a single-relational network analysis algorithm on a multi-relational network by simply ignoring edge labels. This article presents an algebra for mapping multi-relational networks to single-relational networks, thereby exposing them to single-relational network analysis algorithms.
\end{abstract}

\end{frontmatter}

\section{Introduction}

Much of graph and network theory is devoted to understanding and analyzing single-relational networks (also known as directed or undirected unlabeled graphs). A single-relational network is composed of a set of vertices (i.e.~nodes) connected by a set of edges (i.e.~links) which represent relationships of a single type. Such networks are generally defined as $G = (V, E)$, where $V$ is the set of vertices in the network, $E$ is the set of edges in the network, and $E \subseteq (V \times V)$. For example, if $i \in V$ and $j \in V$, then the ordered pair $(i,j) \in E$ represents an edge from vertex $i$ to vertex $j$.\footnote{This article is primarily concerned with directed networks, as opposed to undirected networks, in which an edge is an unordered set of two vertices (e.g.~\{i,j\}). However, the formalisms presented work with undirected networks.} Ignoring the differences in the vertices that edges connect, all edges in $E$ share a single nominal, or categorical, meaning. For example, the meaning of the edges in a single-relational social network may be kinship, friendship, or collaboration, but not all of them together in the same representation as there is no way distinguish what the edges denote. Moreover, single-relational networks may be weighted, where $w : E \rar \mbb{R}$ is a function that maps each edge in $E$ to a real value. Weighted forms are still considered single-relational as all the edges in $E$ have the same meaning; the only difference is the ``degree of meaning" defined by $w$.

The network, as a data structure, can be used to model many real and artificial systems. However, because a network representation of a system can be too complicated to understand directly, many algorithms have been developed to map the network to a lower dimensional space. For a fine review of the many popular network analysis algorithms in use today, refer to \cite{socialanal:wasserman1994,netanal:brandes2005}. Examples of such algorithms, to name a few of the more popularly used algorithms, include the the family of geodesic \cite{close:bavelas1950,short:dijkstra1959,between:freeman1977,eccen:harary1995}, spectral \cite{power:bonacich1987,anatom:brin1998,spectral:chung}, and community detection algorithms \cite{newman-review,girvan-2002,newman-eigen}. 

Most network analysis algorithms have been developed for single-relational networks as opposed to multi-relational networks. A multi-relational network is composed of two or more sets of edges between a set of vertices. A multi-relational network can be defined as $M= (V, \mbb{E})$, where $V$ is the set of vertices in the network, $\mbb{E} = \{E_1, E_2, \ldots, E_m\}$ is a family of edge sets in the network, and any $E_k \subseteq (V \times V) : 1 \leq k \leq m$. Each edge set in $\mathbb{E}$ has a particular nominal, or categorical, interpretation. For example, within the same network $M$, $E_1,E_2 \in \mbb{E}$ may denote ``kinship" and ``coauthorship," respectively.

The multi-relational network is not new. These structures have been used in various disciplines ranging from cognitive science and artificial intelligence \cite{sowa:semantic1991} (e.g.~semantic networks for knowledge representation and reasoning) to social \cite{socialanal:wasserman1994} and scholarly \cite{multigraph:rodriguez2007,semever:bollen2007} modeling. Furthermore, the multi-relational network is  the foundational data structure of the emerging Web of Data \cite{pubsem:lee2001,webinterpret:rodriguez2009}. While a multi-relational network can be used to represent more complicated systems than a single-relational network, unfortunately, there are many fewer multi-relational algorithms than single-relational algorithms. Moreover, there are many more software packages and toolkits to analyze single-relational networks. Multi-relational network analysis algorithms that do exist include random walk \cite{grammar:rodriguez2008}, unique path discovery \cite{discov:lin2004}, community identification \cite{comm:deng2005}, vertex ranking \cite{ranksem:zhuge2003}, and path ranking algorithms \cite{semrank:boan2005}.

The inclusion of multiple relationship types between vertices complicates the design of network algorithms. In the single-relational world, with all edges being ``equal," the executing algorithm  need not be concerned with the meaning of the edge, but only with the existence of an edge. Multi-relational network algorithms, on the other hand, must take this information into account in order to obtain meaningful results. For example, if a multi-relational network contains two edge sets, one denoting kinship ($E_1 \in \mbb{E}$) and the other denoting coauthorship ($E_2 \in \mbb{E}$), then for the purposes of a scholarly centrality algorithm, kinship edges should be ignored. In this simple case, the centrality algorithm can be executed on the single-relational network defined by $G = (V, E_2)$. However, there may exist more complicated semantics that can only be expressed through path combinations and other operations. Thus, isolating single-relational network components of $M$ is not sufficient. As a remedy to this situation, this article presents an algebra for defining abstract paths through a multi-relational network in order to derive a single-relational network representing vertex connectivity according to such path descriptions. From this single-relational representation, all of the known single-relational network algorithms can be applied to yield meaningful results. Thus, the presented algebra provides a means of exposing multi-relational networks to single-relational network analysis algorithms.

\section{Path Algebra Overview}

This section provides and overview of the various constructs of the path algebra and primarily serves as a consolidated reference. The following sections articulate the use of the constructs summarized here.

The purpose of the presented algebra is to transform a multi-relational network into a ``semantically-rich" single-relational network. This is accomplished by manipulating a three-way tensor representation of a multi-relational network. The result of the three-way tensor manipulation yields an adjacency matrix (i.e.~a two-way tensor).\footnote{The term \textit{tensor} has various meanings in mathematics, physics, and computer science. In general, a tensor is a structure which includes and extends the notion of scalar, vector, and matrix. A zero-way tensor is a scalar, a one-way tensor is a vector, a two-way tensor is a matrix, and a three-way tensor is considered a ``cube" of scalars. For a three-way tensor, there are three indices used to denote a particular scalar value in the tensor.} The resultant adjacency matrix represents a ``semantically-rich" single-relational network. The ``semantically-rich" aspect of the resultant adjacency matrix (i.e.~the single-relational network) is determined by the algebraic path description used to manipulate the original three-way tensor (i.e.~the multi-relational network). More formally, the multi-relational path algebra is an algebraic structure that operates on $n \times n$ adjacency matrix ``slices" of a $n \times n \times m$ three-way tensor representation of a multi-relational network in order to generate a $n \times n$ path matrix. The generated $n \times n$ path matrix represents a ``semantically-rich" single-relational network that can be subjected to any of the known single-relational network analysis algorithms. The path algebra is a matrix formulation of the grammar-based random walker framework originally presented in \cite{grammar:rodriguez2008}. However, the algebra generates ``semantically-rich" single-relational networks, as opposed to only executing random walk algorithms in a ``semantically-rich" manner. In other words, the algebra is cleanly separated from the analysis algorithms that are ultimately applied to it. This aspect of the algebra makes it generally useful in many network analysis situations.

The following list itemizes the various elements of the algebra to be discussed in \S \ref{sec:elements}.
\begin{itemize}\setlength{\itemsep}{-1pt}
	\item $\mca{A} \in \{0,1\}^{n \times n \times m}$: a three-way tensor representation of a multi-relational network.\footnote{This article is primarily concerned with boolean tensors. However, note that the presented algebra works with tensors in $\mbb{R}_+^{n \times n \times m}$.}
	\item $\mbf{Z} \in \mbb{R}_+^{n \times n}$: a path matrix derived by means of operations applied to $\mca{A}$.
	\item $\mbf{R}_i \in \{0,1\}^{n \times n}$: a row ``from" path filter.
	\item $\mbf{C}_i \in \{0,1\}^{n \times n}$: a column ``to" path filter.
	\item $\mbf{E}_{i,j} \in \{0,1\}^{n \times n}$: an entry path filter.
	\item $\mbf{I} \in \{0,1\}^{n \times n}$: the identity matrix as a self-loop filter. 
	\item $\mbf{1} \in 1^{n \times n}$: a matrix in which all entries are $1$.
	\item $\mbf{0} \in 0^{n \times n}$: a matrix in which all entries are $0$.
\end{itemize}
The following list itemizes the various operations of the algebra to be discussed in \S \ref{sec:operations}.
\begin{itemize}\setlength{\itemsep}{-1pt}
	\item $\mbf{A} \cdot \mbf{B}$: ordinary matrix multiplication determines the number of $(\mbf{A},\mbf{B})$-paths between vertices.
	\item $\mbf{A}^\top$: matrix transpose inverts path directionality.
	\item $\mbf{A} \circ \mbf{B}$: Hadamard, entry-wise multiplication applies a filter to selectively exclude paths.
	\item $n(\mbf{A})$: \textit{not} generates the complement of a $\{0,1\}^{n \times n}$ matrix.
	\item $c(\mbf{A})$: \textit{clip} generates a $\{0,1\}^{n \times n}$ matrix from a $\mbb{R}_+^{n \times n}$ matrix.
	\item $v^\pm(\mbf{A})$: \textit{vertex} generates a $\{0,1\}^{n \times n}$ matrix from a $\mbb{R}_+^{n \times n}$ matrix, where only certain rows or columns contain non-zero values.
	\item $\lambda\mbf{A}$: scalar multiplication weights the entries of a matrix.
	\item $\mbf{A} + \mbf{B}$: matrix addition merges paths.
\end{itemize}
In short, any abstract path $\sigma$ is a series of operations on $\mca{A}$ that can be generally defined as
\begin{equation*}
	\sigma:  \{0,1\}^{n \times n \times m} \rar \mbb{R}_+^{n \times n}.
\end{equation*}
The resultant $\mbb{R}_+^{n \times n}$ matrix is an adjacency matrix. Thus, the resultant matrix is a single-relational network. This resultant single-relational network can be subjected to single-relational network analysis algorithms while still preserving the semantics of the original multi-relational network.

Throughout the remainder of this article, scholarly examples are provided in order to illustrate the application of the various elements and operations of the path algebra. All of the examples refer to a single scholarly tensor denoted $\mca{A}$. The following list itemizes the tensor ``slices" and their domains and ranges, where $H \subset V$ is the set of all humans, $A \subset V$ is the set of all articles, $J \subset V$ is the set of all journals, $S \subset V$ is the set of all subject categories, and $P \subset V$ is the set of all software programs:
\begin{itemize}\setlength{\itemsep}{-1pt}
	\item $\mca{A}^1$: $\ttt{authored}: H \rar A$
	\item $\mca{A}^2$: $\ttt{authoredBy}: A \rar H$
	\item $\mca{A}^3$: $\ttt{cites}: A \rar A$
	\item $\mca{A}^4$: $\ttt{contains}: J \rar A$
	\item $\mca{A}^5$: $\ttt{category}: J \rar S$
	\item $\mca{A}^6$: $\ttt{developed}: H \rar P$.
\end{itemize} 

\section{Path Algebra Elements}\label{sec:elements}

This section introduces the elements of the algebra and the next section articulates their use in the various operations of the algebra. The elements of the path algebra are structures that are repeatedly used when mapping a multi-relational tensor to a single-relational path matrix. 

\subsection{Three-way Tensor Representation of a Multi-Relational Network}

A single-relational network defined as 
\begin{equation*}
	G = (V, E \subseteq (V \times V))
\end{equation*}
can be represented as the adjacency matrix $\mbf{A}$, where
\begin{equation*}
	\mbf{A}_{i,j} = 
		\begin{cases}
			1 & \text{if } (i,j) \in E \\
			0 & \text{otherwise}
		\end{cases}
\end{equation*}
without loss of information. This adjacency matrix is also known as a two-way tensor because it has two dimensions, each with an order of $n$, where $n = |V|$. Stated another way, $\mbf{A} \in \{0,1\}^{n \times n}$.

A three-way tensor can be used to represent a multi-relational network \cite{tensor:kolda2005}. If
\begin{equation*}
	M = (V, \mbb{E} = \{E_1, E_2, \ldots, E_m \subseteq (V \times V)\})
\end{equation*}
 is a multi-relational network, then
\begin{equation*}
	\mca{A}^{k}_{i,j} = 
		\begin{cases}
			1 & \text{if } (i,j) \in E_k : 1 \leq k \leq m \\
			0 & \text{otherwise}.
		\end{cases}
\end{equation*}
In this formulation, two dimensions have an order of $n$ while the third has an order of $m$, where $m = |\mbb{E}|$. Thus, $\mca{A} \in  \{0,1\}^{n \times n \times m}$ and any adjacency matrix ``slice" $\mca{A}^k \in  \{0,1\}^{n \times n} : 1 \leq k \leq m$. $\mca{A}$ represents the primary structure by which individual adjacency matrices are indexed and composed in order to derive a resultant single-relational path matrix.

\subsection{Path Matrices}

The path algebra operates on $n \times n$ adjacency matrix elements of $\mca{A}$ to construct a ``semantically-rich" path matrix. The simplest path is made up of a single edge type. Thus, $\mca{A}^k$ is a path matrix for vertex-to-vertex paths of length $1$. The meaning of that path is simply defined by the meaning of $\mbb{E}_k$.  When constructing complex paths through $\mca{A}$ (when utilizing multiple edge types in $\mca{A}$), the resulting matrix may have entries with values greater than $1$. Furthermore, in conjunction with the use of $\mbb{R}_+$ scalar multiplication (described in \S \ref{sec:weight}), entries of a path matrix may be non-negative real numbers. Therefore, in general, all path matrices discussed throughout the remainder of this article are matrices in $\mbb{R}_+^{n \times n}$ and are denoted $\mbf{Z}$. In short, the resultant $\mbf{Z}$ matrix can be seen as a positively weighted single-relational network. In other words, $\mbf{Z}$ denotes a single-relational network of the form $G = (V,E,w)$, where $w: E \rar R_+$.

\subsection{Filter Matrices}

Filters are used to ensure that particular paths through $\mca{A}$ are either included or excluded from a path composition. Generally, a filter is a $\{0,1\}^{n \times n}$ matrix. Filters may be derived from intermediate path matrices (described in \S \ref{sec:filter}) or may target a specific vertex that is known \textit{a prior}. A vertex-specific filter is either a row $\mbf{R}_{i} \in \{0,1\}^{n \times n}$, column $\mbf{C}_{i} \in \{0,1\}^{n \times n}$, or entry $\mbf{E}_{i,j} \in \{0,1\}^{n \times n}$ filter.
\begin{enumerate}\setlength{\itemsep}{-1pt}
\item A row filter is denoted $\mbf{R}_i$, where all entries in row $i$ are equal to $1$ and all other entries are equal to $0$. Row filters are useful for allowing only those paths that have an origin of $i$.
\item A column filter is denoted $\mbf{C}_i$, where all entries in column $i$ are equal to $1$ and other entries are equal to $0$. Column filters are useful for allowing only those paths that have a destination of $i$.
\item Entry filters are denoted $\mbf{E}_{i,j}$ and have a $1$ at the $(i,j)$-entry and $0$ elsewhere. Entry filters allow only those paths that have an origin of $i$ or a destination of $j$.
\end{enumerate}
Useful properties of the vertex-specific filters include:
\begin{itemize}\setlength{\itemsep}{-1pt}
	\item $\mbf{R}_{i} = \mbf{C}_{i}^\top$
	\item $\mbf{C}_{i} = \mbf{R}_{i}^\top$
	\item $\mbf{E}_{i,j} = \mbf{E}_{j,i}^\top$.
\end{itemize}

The identity matrix $\mbf{I}$ is useful for allowing or excluding self-loops. Finally, the filter $\mbf{1} \in 1^{n \times n}$ is a matrix in which all entries are equal to $1$, and $\mbf{0} \in 0^{n \times n}$ is a matrix in which all entries are equal to $0$.

\section{Path Algebra Operations}\label{sec:operations}

The previous section defined the common elements of the path algebra. These elements are composed with one another to create a path composition. This section discusses the various operations that are used when composing the aforementioned elements.

\subsection{The Traverse Operation}\label{sec:traverse}

A useful property of the single-relational adjacency matrix $\mbf{A} \in \{0,1\}^{n \times n}$ is that when it is raised to the $t^\text{th}$ power, the entry $\mbf{A}^{(t)}_{i,j}$ is equal to the number of paths of length $t$ that connect vertex $i \in V$ to vertex $j \in V$ \cite{graph:chartrand1977}. This is simple to prove using induction. Given, by definition, that $\mbf{A}^{(1)}_{i,j}$ (i.e.~$\mbf{A}_{i,j}$) represents the number of paths that go from $i$ to $j$ of length $1$ (i.e.~a single edge) and by the rules of ordinary matrix multiplication,
\begin{equation*}
	\mbf{A}_{i,j}^{(t)} =  \sum_{l \in V} \mbf{A}_{i,l}^{(t-1)} \cdot \mbf{A}_{l,j} : t \geq 2.
\end{equation*}

The same mechanism for finding the number of paths of length $t$ in a single-relational network can be used to find the number of semantically meaningful paths through a multi-relational network. For example, suppose $\mca{A}^1$ has the label \ttt{authored}, $\mca{A}^2$ has the label \ttt{authoredBy}, $\mca{A}^3$ has the label \ttt{cites}, and 
\begin{equation*}
	\mbf{Z} = \mca{A}^1 \cdot \mca{A}^3 \cdot \mca{A}^2. 
\end{equation*}
Semantically, $\mbf{Z}_{i,j}$ is the number of paths from vertex $i$ to vertex $j$ such that a path goes from author $i$ to one the articles he or she has authored, from that article to one of the articles it cites, and finally, from that cited article to its author $j$.\footnote{If vertex $i$ is not an author, then such a path composition would yield $0$ paths from $i$ to any $j$. A path composition must respect the domains and ranges of the edge types if a meaningful path matrix is to be generated.} The meaning of $\mbf{Z}$ is $\ttt{hasCited}: H \rar H$ and represents an edge if some author has cited some other author by means of their respective articles (i.e.~an author citation network). This method is analogous to raising an adjacency matrix to the $t^\text{th}$ power, except that, in place of multiplying the same adjacency matrix with itself, a sequence of different adjacency matrix ``slices" in $\mca{A}$ are used to represent typed paths with a compositionally defined meaning.

It is worth noting that ordinary matrix multiplication is not commutative. Thus, for the most part, when $\mbf{A} \neq \mbf{B}$, $\mbf{A} \cdot \mbf{B} \neq \mbf{B} \cdot \mbf{A}$. Given paths through a multi-relational network, this makes intuitive sense. The path from \ttt{authored} to \ttt{cites} is different than the path from \ttt{cites} to \ttt{authored}. In the first case, if the resultant path is seen as a mapping, then $\ttt{authored}: H \rar A$ (i.e.~human to article) and $\ttt{cites}: A \rar A$ (i.e.~article to article). Thus, through composition $\ttt{cites} \circ \ttt{authored}: H \rar A$.\footnote{The symbol $\circ$ is overloaded in this article meaning both function composition and the Hadamard matrix product. The context of the symbol indicates its meaning.} However, in the latter case, composition is not possible as \ttt{cites} has a range of an article and \ttt{authored} has a domain of human.

Finally, any $n \times n$ adjacency matrix element of $\mca{A}$ can be transposed in order to traverse paths in the opposite direction. For example, given  $\mca{A}_1$ and $\mca{A}_2$ identified as \ttt{authored} and \ttt{authoredBy}, respectively, where ${\mca{A}^1}^\top = \mca{A}^2$, then $\mbf{Z} =  \mca{A}^1 \cdot \mca{A}^3 \cdot \mca{A}^2 = \mca{A}^1 \cdot \mca{A}^3 \cdot {\mca{A}^1}^\top$. Thus, inverse edge types can be created using matrix transpose.

\subsection{The Filter Operation}\label{sec:filter}

In many cases, it is important to exclude particular paths when traversing through $\mca{A}$. Various path filters can be defined and applied using the entry-wise Hadamard matrix product denoted $\circ$ \cite{matrix:horn1994}, where
\begin{equation*}
	\mbf{A} \circ \mbf{B} = \left[
	\begin{array}{ccc}
		\mbf{A}_{1,1} \cdot \mbf{B}_{1,1} & \cdots & \mbf{A}_{1,m} \cdot \mbf{B}_{1,m} \\
		\vdots & \ddots & \vdots \\
		\mbf{A}_{n,1} \cdot \mbf{B}_{n,1} & \cdots & \mbf{A}_{n,m} \cdot \mbf{B}_{n,m} \\
	\end{array} \right ] .
\end{equation*}
The following list itemizes various properties of the Hadamard product:
\begin{itemize}\setlength{\itemsep}{-1pt}
	\item $\mbf{A} \circ \mbf{1} = \mbf{A}$
	\item $\mbf{A} \circ \mbf{0} = \mbf{0}$
	\item $\mbf{A} \circ \mbf{B} = \mbf{B} \circ \mbf{A}$
	\item $\mbf{A} \circ (\mbf{B} + \mbf{C}) = (\mbf{A} \circ \mbf{B}) + (\mbf{A} \circ \mbf{C})$
	\item $\mbf{A} \circ \lambda\mbf{B} = \lambda(\mbf{A} \circ \mbf{B})$
	\item $\mbf{A}^\top \circ \mbf{B}^\top = (\mbf{A} \circ \mbf{B})^\top$.
\end{itemize}
If $\mbf{A}$ is a $\{0,1\}^{n \times n}$ matrix, then $\mbf{A} \circ \mbf{A} = \mbf{A}$. For the row, column, and entry filters,
\begin{itemize}\setlength{\itemsep}{-1pt}
	\item $\mbf{R}_{i} \circ \mbf{R}_{j} = \mbf{0} \; : \; i \neq j$
	\item $\mbf{C}_{i} \circ \mbf{C}_{j} = \mbf{0} \; : \; i \neq j$
	\item $\mbf{R}_i \circ \mbf{C}_j = \mbf{E}_{i,j}$.
\end{itemize}
Finally, if $\mbf{Z} \in \mbb{R}_+^{n \times n}$ and $\mbf{Z}$ has a trace of $0$ (i.e.~no self-loops), then $\mbf{Z} \circ \mbf{I} = \mbf{0}$.

The Hadarmard product is used in the path algebra to apply a filter. As stated previously, a typical filter is a $\{0,1\}^{n \times n}$ matrix, where $0$ entries set the corresponding path counts in $\mbf{Z}$ to $0$. The following subsections define and illustrate some useful functions to generate filters.

\subsubsection{The Not Function}

The \textit{not} function is defined as 
\begin{equation*}
	n: \{0,1\}^{n \times n} \rar \{0,1\}^{n \times n}
\end{equation*}
with a function rule of
\begin{equation*}
	n(\mbf{A}) = \mbf{1} - \mbf{A} .
\end{equation*}
In words, the \textit{not} takes a $\{0,1\}^{n \times n}$ matrix and replaces all the $0$s with $1$s and all the $1$s with $0$s. Some evident and useful properties of a \textit{not} filter include
\begin{itemize}\setlength{\itemsep}{-1pt}
	\item $n(n(\mbf{A})) = \mbf{A}$
	\item $\mbf{A} \circ n(\mbf{A}) = \mbf{0}$
	\item $n(\mbf{A}) \circ n(\mbf{A}) = n(\mbf{A})$.
\end{itemize}
Furthermore, if $\mbf{Z} \in \mbb{R}_+^{n \times n}$ and $\mbf{Z}$ has a trace of $0$, then $\mbf{Z} \circ n(\mbf{I}) = \mbf{Z}$.

A \textit{not} function is useful for excluding a set of paths to or from a vertex. For example, when constructing a coauthorship path matrix where $\mca{A}^1$ represents \ttt{authored}, the operation $\mca{A}^1 \cdot {\mca{A}^1}^\top \circ n(\mbf{I})$ will ensure that the \ttt{authored} relationship is taken and then the transpose of \ttt{authored} (i.e.~\ttt{authoredBy}) is taken. However, if only these two operations are applied, then this does not yield a coauthorship matrix, as the traversal returns to the originating vertex (i.e.~vertex $i$ is considered a coauthor of vertex $i$). Thus, the applied \textit{not}-identity filter will remove all paths back to the source vertex, at which point a coauthorship path matrix is generated.

\subsubsection{The Clip Function}

The \textit{clip} function maps an $\mbb{R}_+^{n \times n}$ path matrix to a $\{0,1\}^{n \times n}$ matrix. The function is defined as
\begin{equation*}
	c: \mbb{R}_+^{n \times n} \rar \{0,1\}^{n \times n}
\end{equation*}
with a function rule of
\begin{equation*}
	c(\mbf{Z})_{i,j} = 
		\begin{cases}
			1 & \text{if } \mbf{Z}_{i,j} > 0 \\
			0 & \text{otherwise}.
		\end{cases}
\end{equation*}

The general purpose of \textit{clip} is to take a non-$\{0,1\}^{n \times n}$ path matrix and to ``clip," or normalize, it to a $\{0,1\}^{n \times n}$ matrix. Thus, \textit{clip} creates a filter that can then be applied to a composition to exclude paths.

If $\mbf{A} \in \{0,1\}^{n \times n}$, then evident and useful properties are 
\begin{itemize}\setlength{\itemsep}{-1pt}
	\item $c(\mbf{A}) = \mbf{A}$ 
	\item $c(n(\mbf{A})) = n(c(\mbf{A})) = n(\mbf{A})$.
\end{itemize}

\begin{proposition}
\label{proposition:clip1}
If $\mbf{Y}, \mbf{Z} \in \mbb{R}_+^{n \times n}$, then
\begin{equation*}
	c(\mbf{Y} \circ \mbf{Z}) = c(\mbf{Y}) \circ c(\mbf{Z}).
\end{equation*}
\end{proposition}
\emph{Proof.}
This property can be demonstrated given an entry-wise representation, where
\begin{equation*}
	c(\mbf{Y}_{i,j} \cdot \mbf{Z}_{i,j}) = c(\mbf{Y}_{i,j}) \cdot c(\mbf{Z}_{i,j}).
\end{equation*}
The equality holds for all cases where both entries are $0$, both entries are greater than $0$, and where one entry is $0$ and the other is greater than $0$.
\qed

\begin{proposition}
\label{proposition:clip3}
If $\mbf{A}, \mbf{B} \in \{0,1\}^{n \times n}$, then
\begin{equation*}
	n(\mbf{A} \circ \mbf{B}) = c\left(n(\mbf{A}) + n(\mbf{B})\right)
\end{equation*}
\end{proposition}
\emph{Proof.}
This proposition follows a similar pattern as De Morgan's law for boolean values, where $\neg (P \wedge Q) = \neg P \vee \neg Q$. However, because matrix addition over $\{0,1\}^{n \times n}$ matrices has the potential to yield a value of $2$ if $n(\mbf{A}_{i,j}) = n(\mbf{B}_{i,j}) = 1$, \textit{clip} will ensure that $c(n(\mbf{A}) + c(\mbf{B}))_{i,j} = 1$.\qed

Likewise,
\begin{proposition}
\label{proposition:clip4}
If $\mbf{A}, \mbf{B} \in \{0,1\}^{n \times n}$, then
\begin{equation*}
	n(c(\mbf{A} + \mbf{B})) = n(\mbf{A}) \circ n(\mbf{B})
\end{equation*}
\end{proposition}
\emph{Proof.}
This proposition follows a similar pattern as De Morgan's law for boolean values, where $\neg (P \vee Q) = \neg P \wedge \neg Q$.\qed

To use a scholarly example, it is possible to exclude all coauthorship and self-loop paths from a larger composite. For instance, if, as previously demonstrated, 
\begin{equation*}
	\mca{A}^1 \cdot {\mca{A}^1}^\top \circ n(\mbf{I})
\end{equation*}
defines a coauthorship path matrix and $\mca{A}^3$ denotes \ttt{cites} relations, then
\begin{equation*}
	\mbf{Z} = \underbrace{\left(\mca{A}^1 \cdot \mca{A}^3 \cdot {\mca{A}^1}^\top \right)}_{\text{cites}} \circ 
		\underbrace{n\left(c\left(\mca{A}^1 \cdot {\mca{A}^1}^\top \circ n(\mbf{I})\right)\right)}_{\text{no coauthors}} \circ
		\underbrace{n(\mbf{I})}_{\text{no self}}
\end{equation*}
is a \ttt{hasCited}$'$ path matrix where citing one's coauthors and oneself is not considered a legal citation path. As previously demonstrated in \S \ref{sec:traverse}, the first component (i.e.~cites) generates a \ttt{hasCited} path matrix for all authors citing each other's articles, where coauthorship and self-citation are legal. The second component (i.e.~no coauthors) applies the \textit{not} function to a path matrix generated from a \textit{clip} of a coauthorship path matrix. This excludes coauthors as being legal author citations. Finally, the third component (i.e.~no self) disallows self-loops. The application of the two filter components removes paths that go from an author to his- or herself as well as to his or her respective coauthors.

With the help of the propositions and properties of the various operations of the path algebra, the above composition can be simplified. While the following simplification is lengthy, it utilizes many of the properties and propositions demonstrated hitherto. If $\mbf{X} = \mca{A}^1 \cdot \mca{A}^3 \cdot {\mca{A}^1}^\top$ and $\mbf{Y} = \mca{A}^1 \cdot {\mca{A}^1}^\top$, then
\begin{eqnarray*}
\begin{array}{ccl|l}
	\mbf{Z} &=& \mbf{X} \circ n(c(\mbf{Y} \circ n(\mbf{I}))) \circ n(\mbf{I})  & \\
		    &=& \mbf{X} \circ n(c(\mbf{Y}) \circ c(n(\mbf{I})) \circ n(\mbf{I}) & \text{prop. } \ref{proposition:clip1} \\
		    &=& \mbf{X} \circ n(c(\mbf{Y}) \circ n(\mbf{I})) \circ n(\mbf{I}) & c(n(\mbf{A})) = \mbf{A} \\
		    &=& \mbf{X} \circ c(n(c(\mbf{Y})) + n(n(\mbf{I}))) \circ n(\mbf{I}) & \text{prop. } \ref{proposition:clip3} \\
		    &=& \mbf{X} \circ c(n(c(\mbf{Y})) + \mbf{I}) \circ n(\mbf{I}) & n(n(\mbf{A})) = \mbf{A} \\
		    &=& \mbf{X} \circ c(n(c(\mbf{Y})) + \mbf{I}) \circ c(n(\mbf{I})) & c(n(\mbf{A})) = n(\mbf{A}) \\
		    &=& \mbf{X} \circ c(n(c(\mbf{Y})) + \mbf{I} \circ n(\mbf{I}))) & \text{prop. } \ref{proposition:clip1} \\
		    &=& \mbf{X} \circ c(n(c(\mbf{Y})) \circ n(\mbf{I}) + \mbf{I} \circ n(\mbf{I}))) & \mbf{A} \circ (\mbf{B} + \mbf{C}) = (\mbf{A} \circ \mbf{B}) + (\mbf{A} \circ \mbf{C}) \\
		    &=& \mbf{X} \circ c(n(c(\mbf{Y})) \circ n(\mbf{I})) & \mbf{A} \circ n(\mbf{A}) = \mbf{0} \\
		    &=& \mbf{X} \circ c(n(c(\mbf{Y}))) \circ c(n(\mbf{I})) & \text{prop. } \ref{proposition:clip1} \\
		    &=& \mbf{X} \circ n(c(\mbf{Y})) \circ n(\mbf{I}) & c(n(\mbf{A})) = n(\mbf{A}) .
\end{array}
\end{eqnarray*}
Thus, 
\begin{equation*}
\mbf{Z} = \left(\mca{A}^1 \cdot \mca{A}^3 \cdot {\mca{A}^1}^\top \right) \circ n\left(c\left(\mca{A}^1 \cdot {\mca{A}^1}^\top\right)\right) \circ n(\mbf{I}) .
\end{equation*}
In words, only a single filter disallowing self-loops is necessary to yield the same result. 

\subsubsection{The Vertex Functions}

In many cases, it is important to filter out particular paths from and to a vertex. Two useful functions are $v^-$ and $v^+$, where
\begin{equation*}
	v^-: \mbb{R}_+^{n \times n} \times \mbb{N} \rar \{0,1\}^{n \times n},
\end{equation*}
\begin{equation*}
	v^-(\mbf{Z},p)_{i,j} =
		\begin{cases}
			1 & \text{if } \sum_{l \in V} \mbf{Z}_{i,l} > p \\
			0 & \text{otherwise}
		\end{cases}
\end{equation*}
turns a complete row into an all $1$-row if the sum of row entries is greater than $p$ and
\begin{equation*}
	v^+: \mbb{R}_+^{n \times n} \times \mbb{N} \rar \{0,1\}^{n \times n},
\end{equation*}
\begin{equation*}
	v^+(\mbf{Z}, p)_{i,j} =
		\begin{cases}
			1 & \text{if } \sum_{l \in V} \mbf{Z}_{l,j} > p \\
			0 & \text{otherwise}
		\end{cases}
\end{equation*}
turns a complete column into an all $1$-column if the sum of the column entries is greater than $p$. The function $v^-$ is used to select paths outgoing from particular vertices and $v^+$ is used to select paths incoming to particular vertices. Moreover, by providing $p$, it excludes those vertices with less than $p$ paths outgoing from or incoming to it. For the sake of brevity, when no $p$ is supplied, it is assumed that $p = 0$.

Some useful properties of the vertex filter are
\begin{itemize}\setlength{\itemsep}{-1pt}
	\item $v^-(\mbf{R}_i) = \mbf{R}_i$
	\item $v^+(\mbf{C}_i) = \mbf{C}_i$
	\item $v^+(\mbf{E}_{i,j}) \circ v^-(\mbf{E}_{i,j}) = \mbf{E}_{i,j}$
	\item $v^-(\mbf{Z} \circ \mbf{R}_i) = v^-(\mbf{Z}) \circ \mbf{R}_i$
	\item $v^+(\mbf{Z} \circ \mbf{C}_i) = v^+(\mbf{Z}) \circ \mbf{C}_i$
	\item $v^-(\mbf{Z}, p) = v^+(\mbf{Z}^\top, p)^\top$
	\item $v^+(\mbf{Z}, p) = v^-(\mbf{Z}^\top, p)^\top$.
\end{itemize}

To demonstrate the use of the \textit{vertex} function, consider the multi-relational tensor $\mca{A}$ that includes journals, articles, and subject categories, where $\mca{A}^3$ denotes $\ttt{cites}: A \rar A$, $\mca{A}^4$ denotes $\ttt{contains}: J \rar A$, $\mca{A}^5$ denotes $\ttt{category}: J \rar S$, and vertex $1$ denotes the subject category ``social science." A social science journal citation matrix can be created from $\mca{A}$ in which a path exists between journals if and only if an article contained in a journal cites an article contained in another journal. Furthermore, only those citing and cited articles are considered that are in social science journals. Thus, the social science journal citation path matrix is defined as
\begin{equation*}
\mbf{Z} = \underbrace{\left[v^-\left(\mbf{C}_1 \circ {\mca{A}^5}\right) \circ \mca{A}^4 \right]}_{\text{soc.sci. journal articles}} \cdot
		\mca{A}^3 \cdot
		\underbrace{\left[{\mca{A}^4}^\top \circ  v^+\left(\mbf{R}_1 \circ {\mca{A}^5}^\top\right)\right]}_{\text{articles in soc.sci. journals}}.
\end{equation*} 
First, a \textit{vertex}-created filter is applied to remove all articles that are not contained in social science journals. Next, the articles that these articles cite is determined. Finally, those articles not in social science journals are filtered out using another \textit{vertex}-created filter. Thus, a citation path matrix is generated which only includes social science journals.

Using the various aforementioned \textit{vertex} function properties, the above expression for a social science journal citation path matrix can be simplified, because
\begin{eqnarray*}
\begin{array}{ccl|l}
	v^+\left(\mbf{R}_1 \circ {\mca{A}^5}^\top\right) &=& v^+\left(\left(\mbf{C}_1 \circ \mca{A}^5\right)^\top\right) & \mbf{R}_1 = \mbf{C}_1^\top \\
								   &=& v^-\left(\mbf{C}_1 \circ \mca{A}^5\right)^\top & v^-(\mbf{Z}) = \\ & & & \; \; \;  v^+(\mbf{Z}^\top)^\top .
\end{array}
\end{eqnarray*}
Therefore, given the above and because $\mbf{A}^\top \circ \mbf{B}^\top = (\mbf{A} \circ \mbf{B})^\top$, 
\begin{equation*}
\mbf{Z} = \left[v^-\left(\mbf{C}_1 \circ {\mca{A}^5}\right) \circ \mca{A}^4 \right] \cdot 
	\mca{A}^3 \cdot 
	\left[v^-\left(\mbf{C}_1 \circ {\mca{A}^5}\right) \circ \mca{A}^4 \right]^\top.
\end{equation*}
The above composition reuses the computation for determining which articles are contained in social science journals by simply reversing the directionality of the final component. The two bracketed components represent \ttt{contains} such that the domain is social science journals and the range is articles. The ability to algebraically manipulate path expressions is one of the primary benefits of utilizing an algebraic structure to map a multi-relational network to a single-relational network.

\subsection{The Weight Operation}\label{sec:weight}

Composed paths can be weighted using ordinary matrix scalar multiplication. Given a scalar value of $\lambda \in \mbb{R}$, $\lambda\mbf{Z}$ will weight all the paths in $\mbf{Z}$ by $\lambda$. This operation is useful when merging path matrices in such a way as to make one path matrix more or less significant than another path matrix. The next subsection, \S \ref{sec:merge}, discusses the operation of merging two path matrices and presents an example that includes the weight operation.

\subsection{The Merge Operation}\label{sec:merge}

Ordinary matrix addition can be used to merge two path matrices. For example, consider the multi-relational tensor $\mca{A}$, where $\mca{A}^1$ denotes \ttt{authored} and $\mca{A}^6$ denotes \ttt{developed}, \ttt{authored} maps humans to articles, and \ttt{developed} maps humans to programs. Furthermore, consider a definition of collaboration that includes both the coauthorship of articles and the co-development of software, where article coauthorship is weighted as being slightly more important than co-development. The path matrix
\begin{equation*}
	\mbf{Z} = 0.6\underbrace{\left(\mca{A}^1 \cdot {\mca{A}^1}^\top \circ n(\mbf{I})\right)}_{\text{coauthorship}} \; + \; 0.4\underbrace{\left(\mca{A}^6 \cdot {\mca{A}^6}^\top \circ n(\mbf{I})\right)}_{\text{co-development}}
\end{equation*}
merges the article and software program collaboration path matrices as specified by their respective weights of $0.6$ and $0.4$. The resultant path matrix denotes article and software program collaboration. Finally, using the properties and propositions of the path algebra, a simplification of the previous composition is
\begin{equation*}
	\mbf{Z} = \left[0.6\left(\mca{A}^1 \cdot {\mca{A}^1}^\top\right) + 0.4\left(\mca{A}^6 \cdot {\mca{A}^6}^\top\right)\right] \circ n(\mbf{I}).
\end{equation*}

\section{Network Analysis Applications}

The previous sections presented various elements and operations of the path algebra that can be applied to a multi-relational tensor in $\{0,1\}^{n \times n \times m}$ in order to derive a ``semantically-rich" single-relational path matrix in $\mbb{R}_+^{n \times n}$. The resultant path matrix yields the number of paths from vertex $i$ to vertex $j$ as determined by the operations performed. The path matrix can be considered a weighted single-relational network of the form $G = (V, E, w)$, where $w: E \rar R_+$. Many single-relational network analysis algorithms require either a $\{0,1\}^{n \times n}$ matrix or a $[0,1]^{n \times n}$ weighted or stochastic matrix. The resultant path matrix can be manipulated in various ways (e.g.~normalized out going weight distribution) to yield a matrix that can be appropriately used with the known single-relational network analysis algorithms. This section discusses the connection of a path matrix to a few of the more popular single-relational network analysis algorithms. However, before discussing the single-relational network analysis algorithms, the next subsection discusses the relationship between the path algebra and multi-relational graph query languages.

\subsection{Relationship to Multi-Relational Graph Query Languages}

The presented path algebra has many similarities to graph query languages such as GraphLog \cite{consens:graphlog1990}, Path Query Language (PQL) \cite{graph:leser2005}, and SPARQL Protocol and RDF Query Language (SPARQL) \cite{sparql:prud2004}. All of these languages serve a similar function of querying a graph for data, though they have different levels of expressivity (e.g.~some have mechanisms to include extension functions in the query, to perform regular expressions on the vertex names, and the ability to perform recursion). However, what these languages have in common is the ability to perform graph pattern matching. Graph pattern matching is explained using an example. Suppose the multi-relational network representation $H \subseteq (V \times \Omega \times V)$, where $V$ is the set of vertices and $\Omega$ is the set of edges labels. Given
\begin{equation*}
	Z = \{ ?x \; | \; (?x, \ttt{authoredBy}, \ttt{marko}) \in H \},
\end{equation*}
$Z$ is the set of all articles authored by Marko. In short, $?$-variables are used to bind to particular vertices and must hold for the duration of the query. This is made more salient in the following, more complicated example:
\begin{align*}
	Z =& \{ ?y \; | \; (?x, \ttt{authoredBy}, \ttt{marko}) \in H \\
	    & \;\;\;\;\;\;\;\; \wedge (?x, \ttt{cites}, ?y) \in H \\
	    & \;\;\;\;\;\;\;\; \wedge (\ttt{JOI}, \ttt{contains}, ?y) \in H \\
	    & \;\;\;\;\;\;\;\;\; \wedge \; ?x \neq ?y \; \}.
\end{align*}
In this example, the set $Z$ is the set of all articles that are 
\begin{enumerate}
	\item contained in the Journal of Informetrics (JOI),
	\item cited by Marko's authored articles, and
	\item are not articles authored by Marko. 
\end{enumerate}
In SPARQL, this is represented as

\begin{verbatim}
SELECT ?y WHERE {
  ?x authoredBy marko .
  ?x cites ?y .
  JOI contains ?y .
  FILTER (?x != ?y) }.
\end{verbatim}
In the presented path algebra, this same query is represented as
\begin{equation*}
	\mbf{Z} = c\left( \underbrace{\left[\left(\mbf{C}_2 \circ {\mca{A}^1}^\top \right) \cdot \mca{A}^1 \circ \mbf{I} \right]}_{\text{marko's articles}} \cdot \underbrace{\left[\mca{A}^3 \circ n\left( v^-\left(\mbf{C}_2 \circ {\mca{A}^1}^\top \right)^\top \right) \right]}_{\text{citations to non-marko articles}}  \cdot \underbrace{\left[\mbf{C}_3 \circ {\mca{A}^4}^\top\right]}_{\text{contained in joi}} \right),
\end{equation*}
where vertex $2$ is Marko and vertex $3$ is the Journal of Informetrics. The resulting $\mbf{Z}$ path matrix is a \{0,1\}-matrix, where the row vertices that have a $1$ as an entry are equivalent to those vertices that bound to $?y$ in the related SPARQL query. While the algebraic form may be considered more cumbersome than the SPARQL form, the benefit of using the path algebra is that the query can be simplified through algebraic manipulations, and that it has a convenient implementation using existing linear algebra toolkits. Moreover, a hybrid approach can be utilized that leverages the syntactic convenience of the standard graph query languages and the algebraic properties of the path algebra. That is, as many of the queries used in graph query languages can be specified in the path algebra, it is possible to optimize such queries in the algebra and then convert them back to the query language for execution.

Finally, the path algebra provides more information than a binding of variables to vertices. For instance, in the previous example, without the \textit{clip} function, the resultant $\mbf{Z}$ would return how many paths exist from Marko's articles to those non-Marko cited articles that are contained in the Journal of Informetrics. In this way, some Journal of Informetrics articles may be deemed more ``appropriate" to the query (as there may be more paths to them). The result path matrix can be seen as a weighted single-relational network that can be manipulated further by single-relational network analysis algorithms or used in a larger path expression.

\subsection{Shortest Path Calculation}

The family of geodesic algorithms are based on the calculation of the shortest path between vertices \cite{socialanal:wasserman1994}. Example shortest path metrics include:
\begin{itemize}\setlength{\itemsep}{-1pt}
	\item \textit{eccentricity}: defined for a vertex as the longest shortest path to all other vertices \cite{eccen:harary1995},
	\item \textit{radius}: defined for the network as the smallest eccentricity value for all vertices,
	\item \textit{diameter}: defined for the network as the largest eccentricity value for all vertices,
	\item \textit{closeness}: defined for a vertex as the mean shortest path of a vertex to all other vertices \cite{close:bavelas1950},
	\item \textit{betweenness}: defined for a vertex as the number of shortest paths that a vertex is a part of \cite{between:freeman1977}.
\end{itemize}

A straightforward (though computationally expensive) way to calculate the shortest path between any two vertices $i$ and $j$ in an adjacency matrix $\mbf{A}$ is to continue to raise the adjacency matrix by a power until $\mbf{A}^{(t)}_{i,j} > 0$. The first $t$ where $\mbf{A}^{(t)}_{i,j} > 0$ denotes the length of the shortest path between vertex $i$ and $j$ as $t$. This same principle holds for calculating the shortest path in a path matrix $\mbf{Z}$ and thus, the resulting path matrix generated from a path composition can be used to determine ``semantically-rich" shortest paths between vertices.

\subsection{Diffusing an Energy Vector through $\mca{A}$}

Many network analysis algorithms can be represented as an energy diffusion, where
\begin{equation*}
f:   \mbb{R}_+^{n} \times  \mbb{R}_+^{n \times n} \rar \mbb{R}_+^{n}
\end{equation*}
maps a row-vector of size $n$ (i.e.~``energy vector") and a  matrix of size $n \times n$ to a resultant energy vector of size $n$. Algorithms of this form include eigenvector centrality \cite{power:bonacich1987}, PageRank \cite{anatom:brin1998}, and spreading activation \cite{spread:anderson1983,inform:cohen1987} to name but a few. 

With respect to eigenvector centrality and PageRank, the general form of the algorithm can be represented as a problem in finding the primary eigenvector of an adjacency matrix such that $\pi\mbf{A} = \lambda\pi$, where $\pi \in \mbb{R}^{n}$ and $\pi$ is the energy vector being diffused. Both algorithms can be solved using the ``power method." In one form of the power method, the solution is found by iteratively multiplying $\pi$ by $\mbf{A}$ until $\pi$ converges to a stable set of values as defined by $|| \pi^{(t-1)} - \pi^{(t)} ||_{2} < \epsilon$ for some small $\epsilon \in \mbb{R}_+$. In another form, the problem is framed as finding which $t$-power of $\mbf{A}$ will yield a $\pi$ such that  $\pi\mbf{A}^{(t)} = \lambda\pi$. With respect to spreading activation, the same power method can be used; however, the purpose of the algorithm is not to find the primary eigenvector, but instead to propagate an energy vector some finite number of steps. Moreover, the total energy flow through each vertex at the end of a spreading activation algorithm is usually what is determined.

\subsubsection{The PageRank Path Matrix}

The PageRank algorithm was developed to determine the centrality of web pages in a web citation network \cite{anatom:brin1998} and since, has been used as a general network centrality algorithm for various other types of networks including bibliographic \cite{journalstatus:bollen2006,findgem:chen2006}, social \cite{coauth:liu2005}, and word networks \cite{pagesem:mihalcea2004}. The web citation network can be defined as $G = (V, E)$, where $V$ is a set of web pages and $E \subseteq (V \times V)$ is the set of directed citations between web pages (i.e.~\ttt{href}). The interesting aspect of the PageRank algorithm is that it ``distorts" the original network $G$ by overlaying a ``teleportation" network in which every vertex is connected to every other vertex by some weight defined by $\delta \in (0,1]$. The inclusion of the teleportation network ensures that the resulting hybrid network is strongly connected\footnote{Strongly connected means that there exists a path from every vertex to every other vertex. In the language of Markov chains, the network is irreducible and recurrent.} and thus, the resultant primary eigenvector of the network is a positive real-valued vector.

In the matrix form of PageRank, there exist two adjacency matrices in $[0,1]^{n \times n}$ denoted
\begin{equation*}
	\mca{P}^1_{i,j} = 
		\begin{cases}
			\frac{1}{\Gamma(i)} & \text{if } (i,j) \in E \\
			0 & \text{otherwise}.
		\end{cases}
\end{equation*}
and
\begin{equation*}
	\mca{P}^2_{i,j} = \frac{1}{|V|},
\end{equation*}
where $\Gamma(i)$ is the out degree of vertex $i$. $\mca{P}^1$ is a row-stochastic adjacency matrix and $\mca{P}^2$ is a fully connected adjacency matrix known as the teleportation matrix. The purpose of PageRank is to identify the primary eigenvector of a merged, weighted path matrix of the form
\begin{equation*}
\mbf{Z} = \delta \mca{P}^1+ (1-\delta) \mca{P}^2.
\end{equation*}
$\mbf{Z}$ is guaranteed to be a strongly connected single-relational path matrix because there is some probability (defined by $1 - \delta$) that every vertex is reachable by every other vertex.

\subsubsection{Constrained Spreading Activation}

The concept of spreading activation was made popular by cognitive scientists and the connectionist approach to artificial intelligence \cite{spread:anderson1983,spread:collins1975,rumelhart:conn1993}, where the application scenario involves diffusing an energy vector through an artificial neural network in order to simulate the neural process of a spreading activation potential in the brain. Spreading activation can be computed in a manner analogous to the power method of determining the primary eigenvector of an adjacency matrix. However, spreading activation does not attempt to find a stationary energy distribution in the network and moreover, usually includes a decay function or activation/threshold function that can yield a resultant energy vector whose sum is different than the initial energy vector.

The neural network models of connectionism deal with weighted, single-relational networks. Spreading activation was later generalized to support multi-relational networks \cite{inform:cohen1987}. While there are many forms of spreading activation on single-relational networks and likewise, many forms on multi-relational networks, in general, a spreading activation algorithm on a multi-relational network is called constrained spreading activation as not all edges to and from a vertex are respected equally. Constrained spreading activation has been used for information retrieval on the web \cite{inform:cohen1987,applic:crestani1997,search:crestani2000}, semantic vertex ranking \cite{grammar:rodriguez2008,socialgrammar:rodriguez2007}, and collaborative filtering \cite{griffen:spread2006}. The path algebra provides a means by which to separate the spreading activation algorithm from the data structure being computed on. That is, the \textit{constrained} aspect of the algorithm is defined by the path algebra and the \textit{spreading activation} aspect of the algorithm is defined by standard, single-relational spreading activation. 

\subsection{Mixing Patterns}

Given a network and a scalar or categorical property value for each vertex in the network, it is possible to determine whether the network is assortative or disassortative with respect to that property \cite{newman:mixpatt2003}. Example scalar properties for journal vertices in a scholarly network include impact factor ranking, years in press, cost for subscription, etc. Example categorical properties include subject category, editor in chief, publisher, etc. A network is assortative with respect to a particular vertex property when vertices of like property values tend to be connected to one another. Colloquially, assortativity can be  understood by the phrase ``birds of a feather flock together," where the ``feather" is the property binding individuals \cite{homophily:cook2001}. On the other hand, a network is disassortative with respect to a particular vertex property when vertices of unlike property values tend to be connected to one another. Colloquially, disassortativity can be understood by the phrase ``opposites attract."

There are two primary components needed when calculating the assortativity of a network. The first component is the network and the second component is a set of property values for each vertex in the network. A popular publication defining the assortative mixing for scalar properties uses the parametric Pearson correlation of two vectors \cite{newman:assort}.\footnote{Note that scalar value distributions may not be normally distributed and thus, in such cases, a non-parametric correlation such as the Spearman $\rho$ may be the more useful correlation coefficient.} One vector is the scalar value of the vertex property for the vertices on the tail of all edges. The other vector is the scalar value of the vertex property for the vertices on the head of all the edges. Thus, the length of both vectors is $|E|$ (i.e.~the total number of edges in the single-relational network).  Formally, the correlation is defined as
\begin{equation*}
r = \frac{|E| \sum_{i} j_i k_i -  \sum_i j_i \sum_i  k_i}{\sqrt{\left[|E| \sum_i j^2_i - \left(\sum_i j_i\right)^2\right]\left[|E| \sum_i k^2_i -\left(\sum_i k_i\right)^2\right]}},
\end{equation*}
where $j_i$ is the scalar value of the vertex on the tail of edge $i$, and $k_i$ is the scalar value of the vertex on the head of edge $i$. The correlation coefficient $r$ is in $[-1,1]$, where $-1$ represents a fully disassortative network, $0$ represents an uncorrelated network, and $1$ represents a fully assortative network. On the other hand, for categorical properties, the equation
\begin{equation*}
r = \frac{\sum_{a} e_{aa} - \sum_a i_a j_a}{1 - \sum_a i_a j_a}
\end{equation*}
yields a value in $[-1,1]$ as well, where $e_{aa}$ is the number of edges in the network that have property value $a$ on both ends, $i_a$ is the number of edges in the network that have property value $a$ on their tail vertex, and $j_a$ is the number of edges that have property value $a$ on their head vertex \cite{newman:mixpatt2003}.

Given a path matrix, assortativity is calculated on paths, not on edges, where there may be many paths (possibly weighted) between any two vertices.\footnote{It is important to note that with multi-relational networks, vertex property values can be encoded in the network itself. For instance, given the scholarly network example of previous, the subject category ``social science" is a vertex adjacent to a set of journal vertices according to the relation $\ttt{category}: J \rar S$. More generally, vertex property values may be determined through path composition.} Thus, a weighted correlation is required \cite{weightedcorr:bland1995}. Let $j_i$ denote the scalar property value of the vertex on the tail of path $i$ and $k_i$ denote the scalar property value of the vertex on the head of path $i$. The previous scalar assortativity equation can be generalized such that if $z_i$ is the fraction of path weight in $\mbf{Z}$ for path $i$, then
\begin{equation*}
r = \frac{\text{cov}_{j,k}}{\sqrt{\text{cov}_{j,j}\text{cov}_{k,k}}},
\end{equation*}
where
\begin{equation*}
\text{cov}_{j,k} = \frac{1}{\sum_i z_i} \left[ \sum_i z_i \left(j_i - \frac{\sum_i z_i j_i}{\sum_i z_i}\right) \left(k_i - \frac{\sum_i z_i k_i}{\sum_i z_i}\right) \right].
\end{equation*}
 Similarly, for categorical vertex properties,
\begin{equation*}
r = \frac{\sum_{a} e_{aa} - \sum_a i_a j_a}{1 - \sum_a i_a j_a}
\end{equation*}
where $e_{aa}$ is the total path weight of paths that have tail and head vertices with a property value of $a$, $i_a$ is the total path weight of all paths that have a tail vertex with a property value of $a$, and $j_a$ is the total path weight of all paths that have a head vertex with a property value of $a$.

\section{Conclusion}

The number of algorithms and toolkits for single-relational networks far exceeds those currently available for multi-relational networks. However, with the rising popularity of the multi-relational network, as made evident by the Web of Data initiative and the multi-relational RDF data structure \cite{rdfintro:miller1998}, there is a need for methods that port the known single-relational network analysis algorithms over to these multi-relational domains. A convenient method to do so is a path algebra. Path algebras have been used extensively for analyzing paths in single-relational networks \cite{carre:graphs1979,path:manger2004} and the application to multi-relational networks can prove useful. The multi-relational path algebra presented in this article operates on an $n \times n \times m$ tensor representation of a multi-relational network.  By means of a series of operations on two-way ``slices" of this tensor, a ``semantically-rich" $n \times n$ single-relational path matrix can be derived. The resulting path matrix represents a single-relational network. This single-relational network may then be subjected to any of the known single-relational network analysis algorithms. Thus, the presented path algebra can be used to expose multi-relational networks to single-relational network analysis algorithms.

\end{document}